Cécile FAUVELOT[1,2], Francesca BERTOZZI[1], Federica COSTANTINI[1], Laura AIROLDI[1], Marco ABBIATI[1]

# Lower genetic diversity in the limpet *Patella caerulea* on urban coastal structures compared to natural rocky habitats

[1] Centro Interdipartimentale di Ricerca per le Scienze Ambientali and Dipartimento di Biologia Evoluzionistica Sperimentale, University of Bologna, Via S. Alberto 163, I – 48100 Ravenna, Italy

[2] Institut de Recherche pour le Développement (IRD) – UR128, Centre de Biologie et d'Ecologie Tropicale et Méditerranéenne, Université de Perpignan, 52 Av. Paul Alduy, F - 66860 Perpignan cedex, France.

Corresponding author:

Dr. Cécile Fauvelot
IRD – UR128
Centre de Biologie et d'Ecologie Tropicale et Méditerranéenne
Université de Perpignan
52 Av. Paul Alduy
F - 66860 Perpignan cedex, France
Tel : +33 4 68 66 20 55
Fax: +33 4 68 50 36 86
E-mail: cecile.fauvelot@univ-perp.fr

**ABSTRACT**

Human-made structures are increasingly found in marine coastal habitats. The aim of the present study was to explore whether urban coastal structures can affect the genetic variation of hard-bottom species. We conducted a population genetic analysis on the limpet *Patella caerulea* sampled in both natural and artificial habitats along the Adriatic coast. Five microsatellite loci were used to test for differences in genetic diversity and structure among samples. Three microsatellite loci showed strong Hardy-Weinberg disequilibrium likely linked with the presence of null alleles. Genetic diversity was significantly higher in natural habitat than in artificial habitat. A weak but significant differentiation over all limpet samples was observed, but not related to the type of habitat. While the exact causes of the differences in genetic diversity deserve further investigation, these results clearly point that the expansion of urban structures can lead to genetic diversity loss at regional scales.

Key words: *Patella caerulea*; genetic diversity; microsatellites; Adriatic Sea; null alleles; coastal urbanization; artificial rocky habitats.

## INTRODUCTION

Human-made structures (such as sea walls, breakwaters, groynes, dykes and other rock armoured urban structures) are increasingly built in marine coastal habitats for a variety of purposes. A recent review of the status of European coastlines (Airoldi and Beck 2007) has shown that, nowadays, 22000 $km^2$ of the European coastal zone are covered in concrete or asphalt, and that urban artificial surfaces have increased by nearly 1900 $km^2$ between 1990 and 2000 alone. Similar examples occur in other parts of the world - e.g. California (Davis et al. 2002), Australia (Connell 2001) and Japan (Koike 1996) - where hundreds of kilometres of coasts are hardened to some extent.

In most instances, artificial hard structures are built in areas which otherwise have soft sediment habitats (e.g. breakwaters on sandy shores, Figure 1). These artificial substrata may alter native soft-bottom assemblages (Martin et al. 2005) and promote the establishment of non-native hard-bottom species (Bulleri and Airoldi 2005; Moschella et al. 2005) creating unnatural changes in species composition, abundance and diversity (Airoldi et al. 2005a,b; Bulleri 2005). This suggests that the expansion of urban structures may be one of the major drivers of biotic homogenization (McKinney 2006). Numerous benthic organisms dwelling on artificial structures rely on a pelagic larval phase to disperse and colonize new habitats. As a consequence, the introduction of artificial hard structure may provide new substrates for invasive species but may also generate novel ecological corridors for native hard bottom species by increasing the connectivity among isolated (e.g. by stretches of sandy habitats) and differentially adapted populations. While the spread of aquatic invasive species through human mediated introductions has received wide consideration (see Roman and Darling 2007 for a review), surprisingly, limited attention has been paid to the possible role of marine urban coastal structures in connecting discrete populations of native hard bottom species (Dethier et al.

2003) and in locally modifying genetic diversity in populations inhabiting artificial structures, recently made available for colonization.

Genetic diversity within a population can affect the productivity, growth and stability, as well as inter specific interaction within community, and ecosystem-level processes (Hughes et al. 2008). Importance of genetic diversity in adaptation processes is well documented and crucial for species survival in highly variable environment or those subject to rapid anthropogenic changes (see Reusch et al. 2005 for an example). Moreover, recent studies have shown that increasing genetic diversity within species can have positive effects on coexistence of competing species (Vellend 2006).

The aim of the present study is to explore whether urban coastal structures can affect the genetic diversity and structure of hard-bottom species. We tested this hypothesis along the coastlines of the Adriatic Sea. In this region, extensive and uncontrolled urbanization during the past century has caused the proliferation of hundreds of kilometers of hard coastal artificial structures, which are now particularly abundant along the Italian sandy shores (Figure 1, see below “Study area and species”). We focussed on the limpet *Patella caerulea,* one of the most common and numerically abundant intertidal species found on both artificial structures and natural rocky shores in this region. Limpets have a key role in structuring intertidal and shallow subtidal rocky shores assemblages, and factors affecting their distribution can cause significant changes in these systems (Jenkins et al 2005). We used a comparative spatial framework (artificial hard structures *versus* natural rocky shores) and microsatellite molecular markers to examine genetic diversity and structure of samples of the limpet *Patella caerulea*. Based on a hierarchical sampling design replicated in several locations (Figure 2), we specifically tested 1) for possible differences in the genetic structuring of populations between natural shores and artificial structures, and 2) whether the genetic diversity of populations on artificial structures was reduced compared to populations on natural reefs, as could be expected from the recent founding of artificial substrates, or alternatively enhanced through the increased number of human mediated introduction vectors.

## MATERIAL AND METHODS

### Study area and species

Within the Adriatic Sea, the Italian coast consists of a sandy flat coastal system almost uninterrupted, in contrast to the prevailingly rocky shores of the Balkans. Along the Italian shoreline, natural hard-bottom habitats are scarce and represented by isolated rocky promontories (from North East to South: Sistiana/Miramare, Gabicce, Conero and Gargano, see Figure 2). Human-made structures (mainly rock-armoured breakwaters, but also groynes, seawalls and harbour jetties) have proliferated on these sandy coasts along hundreds of kilometers of coast (Figure 1 and 2), with most coastal defense structures built since the 80ies (Cencini 1998). The artificial structures included in the present study were offshore detached breakwaters, built with large blocks of quarried rock (mainly limestone), and set on shallow sediments. The assemblages and main ecological characteristics of urban coastal structures in this regions are described in Bacchiocchi and Airoldi (2003), Airoldi et al. (2005b), Bulleri and Airoldi (2005). Information on the geomorphology, hydrology and environmental characteristics of the Adriatic Sea can be found in Poulain (2001).

The limpet *Patella caerulea,* is common along the Adriatic coastline. It is patchily distributed and tends to be up to three times more abundant on artificial structures than on natural rocky shores, reaching on some structures peak densities above 600 ind.m$^{-2}$ (Airoldi et al. unpublished data). *P. caerulea* is a sedentary species, and colonises new isolated habitats, such as those provided by artificial urban structures, by means of dispersing larvae. Its spawning period is ranging from September to April, with a peak in mid winter (Bacci and Sella 1970; Airoldi et al. unpublished data). Limpets are long-lived broadcast spawners. After a brief embryonic period, offspring hatch as free-swimming trochophores (Buckland-Nicks et al. 2002). Little information is available on life history and effective dispersal of *P. caerulea*. Larval duration and behaviour is not known and in the closely related species *Patella vulgata* the larval period is up to 12 days long with a pre-competency period of 4 days (Dodd 1957).

**Sampling**

Limpets were sampled at mid intertidal levels (10 to 30 cm above Mean Low Water Level). Sampling was conducted repetitively in different locations following a hierarchical design (Figure 2). At each of the selected locations where natural rocky shores occurred (Trieste, Ortona, Gallipoli and Split) limpets were collected from 2 natural sites approximately 2 km apart. We included the two sampling sites of Split and Gallipoli in order to acquire a genetic picture of *P. caerulea* populations in prevailing natural rocky shores. In Trieste and Ortona, where both artificial and natural rocky substrata occur, limpets were also sampled on artificial structures at 2 sites, to provide a comparative framework to test for genetic differences between artificial and natural substrata. Artificial structures were few 100m apart from natural rocky coasts, and were spaced about 2 km apart, similarly to the natural sites. One additional sampling was carried out following the same design in Cesenatico where only artificial structures are present and the closest natural rocky shores is > 40 km apart. Sampling was carried out during the summers 2002 (for Cesenatico, Trieste and Ortona) and 2004 (for Gallipoli and Split). For each sampling site, either on natural rocks or on artificial structures, 21 to 50 specimens of *P. caerulea* were randomly collected, for a total of 549 limpets. Specimens collected were generally larger than 15 mm, thus not including juveniles. Live specimens were transported to the laboratory, foot muscle were cut and stored at -80°C until processing.

## Microsatellite isolation and genotyping

A dinucleotide-enriched partial genomic library has been constructed using the FIASCO protocol (Zane et al. 2002). Genomic DNA was extracted from frozen foot muscle tissue of a single individual using the CTAB extraction procedure (Winnepenninckx et al. 1993) as described in Costantini et al. (2007). Following extraction, DNA was simultaneously digested with *Mse*I, ligated to *Mse*I-adaptors (5'- TACTCAGGACTCAC - 3'/ 5' - GACGATGAGTCCTGAG - 3') and amplified with *Mse*I adaptor specific primers (5'-GATGAGTCCTGAGTAA(CATG)-3': hereafter referred as *MseI*-N). The 20µl PCR reaction contained 1x PCR buffer (Promega), 1.5mM $MgCl_2$, 120ng primer *Mse*I-N, 0.2mM of each dNTP, 0.4 units Taq polymerase (Promega) and 5 µl of a 1/10 dilution of the digested-ligated product. PCRs were carried out in a GeneAmp® PCR System 2700 (Applied Biosystems): 94 °C 30 s, 53 °C 1 min, 72°C 1 min for 20 cycles. Amplified DNA was hybridised with a biotinylated probe $(AC)_{17}$ (denaturation of 3 min at 95°C followed by a 15 min annealing at room temperature), selectively captured using streptavidine-coated beads (Roche) and separated by a magnetic field. DNA was eluted from the beads-probe with TE 1x buffer (Tris-HCl 10 mM, EDTA 1mM, pH 8) at 95 C° for 5 min, precipitated with sodium acetate and ethanol, re-amplified by 30 cycles of PCR using the *Mse*I-N primer under the conditions described above, and cloned using the TOPO-TA cloning kit (Invitrogen) following the manufacturer's protocol. Recombinant clones were screened by PCR amplification with M13 forward-reverse primers and sequenced using the BigDye Terminator Cycle Sequencing kit (Applied Biosystem) and resolved on a ABI 310 Genetic Analyser (Applied Biosystem).

About 200 colonies were screened and sequenced for the presence of simple sequence repeats. Analyses revealed the occurrence of repeats in 55 clones. After excluding loci with too short flanking regions, primers for more than 40 loci were designed using the Primer 3 program (Rozen and Skaletsky 1998). Primer pairs were then optimised for PCR amplification testing over a range of annealing temperatures and $MgCl_2$ concentrations. Excluding loci that failed to amplify or resulted in monomorphic patterns, five polymorphic dinucleotide microsatellite remaining loci were reliably amplified in all tested individuals (Table 1).

For all collected *P. caerulea*, DNA was isolated as described above and samples were screened for variation at the five loci newly isolated and optimized. The 20μl PCR reaction contained about 50ng of genomic DNA, 1.0-1.5mM $MgCl_2$ (Table 1), 0.5μM of each primer, 0.2mM of each dNTP, 10mM Tris-HCl (pH 9), 50mM KCl, 0.1% Triton X-100 and 1U of Taq polymerase (Promega). PCR reactions were performed on a GeneAMP PCR System 2700 (Applied Biosystems): denaturation for 3 min at 94°C, followed by 30 cycles of 30s at 94°C, 30s at 55°C, and 30s at 72°C, and a final holding at 72°C for 5 min. Amplified fragments were run on an ABI310 automated Genetic Analyser (Applied Biosystems), using forward primers 5'-labelled with 6-FAM, HEX or TAMRA (MGW Biotech) and the ROX HD400 (Applied Biosystems) as internal standard. Genotyping of individuals was performed by allele sizing using the GeneScan Analysis Software v. 2.02 (Applied Biosystems).

**Data analysis**

Observed heterozygosity ($H_O$) and unbiased gene diversity ($H_S$, Nei 1987) were calculated within each population for each locus and overall loci in Genetix (Belkhir et al. 2004), and multilocus allelic richness (*Ar*, El Mousadik and Petit 1996) was computed in Fstat v.2.9.3 (Goudet 1995, 2001). Significant differences in genetic diversity ($H_O$, $H_S$, and *Ar*) among groups of samples (related to natural *versus* artificial habitats) were tested using a permutation procedure (10000 iterations) in FSTAT. Linkage disequilibrium between loci, and deviations from Hardy-Weinberg (HW) expectations were tested using Fisher's exact tests based on Markov chain procedures in GENEPOP v.3.4 (Raymond and Rousset 1995) as implemented for online uses (http://genepop.curtin.edu.au/). Significance levels for multiple comparisons of loci across samples were adjusted using a standard Bonferroni correction (Rice 1989). The presence of null alleles was examined by estimating null allele frequencies for each locus and sample following the Expectation Maximization (EM) algorithm of Dempster et al. (1977) using FREENA (Chapuis and Estoup 2007).

In order to reveal the presence (if any) of genetic bottleneck signatures in the 14 samples populations, we used the *M* ratio of number of alleles *k* divided by the allelic size range *r*, averaged across all loci in each sample (Garza and Williamson 2001). This ratio calculated over all loci for each sample using the program M_P_VAL (Garza and Williamson 2001) is intended to quantify gaps in the allele size frequency distribution resulting from loss of alleles through bottlenecking. The observed values of *M* average over all loci were then compared to the equilibrium distribution of *M* simulated according to the method described in Garza and Williamson (2001), and given values of theta, *ps* (proportion of one-step mutations) and $\Delta g$ (average size of non one-step mutations) set to 2, 0.8 and 3.5 respectively (Garza and Williamson 2001). If the observed value of *M* is lower than the critical value of *M*, *Mc*, (defined such that only 5% of the simulation values fall below), it is taken as evidence that the sample is from a population that had experienced a recent bottleneck/founding.

Genetic divergences among samples were estimated using the $F_{ST}$ estimates of Weir (1996) and following the so-called ENA method described in Chapuis and Estoup (2007) since the presence of null alleles was found (see results). The null allele frequencies are estimated based on Hardy-Weinberg equilibrium, the genotypes are adjusted based on the null allele frequencies, and the ENA method provides unbiased $F_{ST}$ estimates based on the adjusted data set. These calculations were conducted using FreeNA (Chapuis and Estoup 2007). Owing to deviation from Hardy-Weinberg equilibrium, genotypic differentiation among samples was tested with an exact test (Markov chain parameters: 1000 dememorizations, followed by 1000 batches of 1000 iterations per batch), and the *P*-value of the log-likelihood (G) based exact test (Goudet et al. 1996) was estimated in Genepop. Significant threshold values were adjusted with a sequential Bonferroni correction (Rice 1989) that corrects for sampling error associated with multiple tests.

In order to examine the partition of the genetic variance among limpet samples based on the type of habitat to further test the impact of artificial reefs in possibly increasing populations' connectivity, an analysis of molecular variance (AMOVA) (Excoffier et al. 1992) implemented in ARLEQUIN version 3.1 (Excoffier et al. 2005) was conducted on the original dataset (i.e. not adjusted for null alleles).

**RESULTS**

For the five reliably amplified and analyzed microsatellite loci, number of alleles ranged from 13 for *Pc11* to 30 for *Pc38*. Over all microsatellite loci, highly significant multilocus deviations from HW proportions were observed in all 14 samples (Table 2). At single loci, nearly all comparisons (66 out of 70) showed heterozygote deficiencies, from which 46 showed significant heterozygote deficiencies after Bonferroni corrections. In particular, three of the five scored microsatellites (*Pc15*, *Pc36* and *Pc73*) showed strong heterozygote deficiencies in all 14 samples (i.e. equally affecting all samples), suggesting the presence of null alleles, while the two others (*Pc11* and *Pc38*) were in HW equilibrium in nearly all samples (though $H_O < H_S$ in nearly all cases). Assuming HWE, estimated null allele frequencies (*R*) ranged among loci from 0 to 0.66 (Table 2). The number of expected null homozygotes within samples based on HW equilibrium ($N*R^2$) was significantly higher than the average number of observed null homozygotes for *Pc15*, *Pc38* and *Pc73* (paired t-test, $P = 0.004$, $P = 0.026$ and $P < 0.001$ respectively), suggesting that although null alleles are present in the dataset, they were overestimated.

Over all loci, allelic richness within samples based on a minimum sample size of 12 diploid individuals (i.e. the number of genotypes at *Pc36* in Split1) ranged from 8.09 in Tri1A to 9.42 in Ort1N (Table 2, Figure 3A) and gene diversity from 0.76 in Tri1A to 0.86 in Spl2N (Table 2). Allelic richness (Figure 3B) and gene diversity were both significantly higher in natural habitat (average and standard deviation: $Ar = 8.98 \pm 0.29$; $H_S = 0.836 \pm 0.01$) than in artificial habitat (average $Ar = 8.40 \pm 0.25$; $H_S = 0.805 \pm 0.02$; p-values associated with the permutation procedure: $P = 0.0008$ and $P = 0.0004$, respectively). Among the four sampled sites for which both natural and artificial habitats were sampled (Tri1, Tri2, Ort1 and Ort2), allelic richness based on a minimum sample size of 22 individuals (Figure 3C) and gene diversity were both significantly higher in natural ($Ar = 10.97 \pm 0.55$, $H_S = 0.838 \pm 0.01$) than in artificial habitats ($Ar = 10.16 \pm 0.06$, $H_S = 0.805 \pm 0.03$; permutation procedure, $P = 0.014$ for both tests). Conducting the same comparisons, but only with *Pc11* and *Pc38* (the two loci in HWE), results are similar with both allelic richness (based on 21 individuals) and gene diversity being significantly higher in natural habitat (average and standard deviation: $Ar = 11.77 \pm 0.99$; $H_S = 0.846 \pm 0.02$) than in artificial habitat (average $Ar = 10.71 \pm 0.27$; $H_S = 0.824 \pm 0.01$; p-values associated with the permutation procedure: $P = 0.006$ and $P = 0.005$, respectively).

None of the sampled populations experienced a recent bottleneck since none of the *M* ratios for individual sites fell under the lower 5% of the distribution of simulated *M* values. The *M* ratio for the samples ranged from 0.67 for Gal1 to 0.87 for Gal2 (*Mc* = 0.57 for a sample size of 40 individuals, 5 loci and given the three parameters used for the simulations).

Over all 14 samples, the multilocus $F_{ST}$ estimate was low (0.0094) though the overall genotypic differentiation was significant ($P < 0.0001$). Pairwise $F_{ST}$ estimates among the 14 samples ranged from 0 to 0.032 (between Tri1A and Gal1N) and significant genotypic differentiations among samples after sequential Bonferroni corrections were found in 16 out of 91 comparisons (Table 3). The sample Tri1A appeared the most differentiated from all other samples with pairwise $F_{ST}$ estimates of 0.013-0.032. Over all the five sampling locations (i.e. when pooling samples according to the sampling location), multilocus $F_{ST}$ estimate decreased to 0.0051 ($P < 0.0001$).

The AMOVAs conducted on the dataset not adjusted for the presence of null alleles showed a weak but significant differentiation among limpet samples ($F_{ST} = 0.016$, $P < 0.001$). The nuclear variance attributed to the type of habitat was not significant either across all samples (Variance component = 0.003, $P = 0.152$) or including only the sites where both natural and artificial sites were sampled (Variance component = -0.007, $P = 0.972$).

## DISCUSSION

Urban coastal structures offer suitable substrata for the colonization of *P. caerulea*, up to the point that at some sites (e.g. Cesenatico) this limpet is three times more abundant on recently built urban coastal structures than on nearby natural rocky shores (Airoldi et al., unpublished data). At the same time, the present results showed that the genetic diversity within populations of *P. caerulea* is significantly smaller on artificial structures than on natural reefs. No evidence of genetic differentiation between artificial and natural substrates was found at the five neutral molecular markers studied, and a subtle genetic structure was found over all Adriatic samples.

*Hardy-Weinberg equilibrium and null alleles*

Nearly all loci at all sites showed heterozygote deficiencies, with extremely strong deficiencies observed at three loci (*Pc15*, *Pc36* and *Pc73*). Null alleles were present at these three loci, as revealed by the occurrence of null homozygotes (i.e. non amplifying individuals at some loci), but their occurrence appeared overestimated assuming HW equilibrium within samples. Though likely overestimated null allele frequencies in our study are high (up to 0.66), they fall in the range of null allele frequencies presented in Dakin and Avise (2004) based on 74 microsatellite loci from a wide range of organisms, notably with large effective population sizes (Chapuis and Estoup 2007). Although null alleles lead to underestimated genetic diversity within samples (Paetkau and Strobeck 1995), it is a minor source of error in estimating heterozygosity excess for the detection of bottlenecks (Cornuet and Luikart 1996) and in parental assessments (Dakin and Avise 2004). Moreover, though estimates of differentiation and the probability of detecting genetic differences among populations both diminished when locus heterozygosities are high and data corrected for null alleles (O'Reilly et al. 2004; Peijnenburg et al. 2006; present results), in the presence of null alleles, $F_{ST}$ estimates are unbiased in the absence of population structure (Chapuis and Estoup 2007). This is likely the case in our study since we found that adjusting our data set according to the presence of null alleles did not alter our conclusions regarding the low levels of genetic structure (overall mutlilocus $F_{ST}$ estimated from adjusted data set in FreeNA = 0.009; overall mutlilocus $F_{ST}$ estimated from original data set in ARLEQUIN = 0.016).

Heterozygote deficiencies have already been observed in *P. caerulea* populations analysed using allozymes with no null homozygotes observed (Mauro et al. 2001). Consistency between microsatellite and allozyme data suggest that heterozygote deficiencies may be partially explained by a Wahlund effect (i.e. fine scale genetic patchiness), a common feature in limpets, as well as in other marine invertebrates (e.g. Côrte-Real et al. 1996; Costantini et al. 2007; Hurst and Skibinski 1995; Johnson and Black 1984; Pérez et al. 2007). Such localised genetic heterogeneity could result from spatial or temporal heterogeneity in the genetic composition of recruits, or from post-settlement selection (Johnson and Black 1984).

### *Genetic structure of Adriatic* P. caerulea *populations*

*P. caerulea* population genetic analysis at five neutral molecular markers revealed a weak but significant structure in the Adriatic Sea, mostly associated with the distinctiveness of one Trieste sample (Tri1A). The mean multilocus $F_{ST}$ estimate was very low over all the 14 sample sites (0.009, $P < 0.0001$), comparable to what was found by Mauro et al. (2001) using allozymes across the same region and similar spatial scales (0.007, $P > 0.05$). Also, a lack of significant differentiation of the Trieste sample from Sicily samples was observed using allozymes (Mauro et al. 2001). Therefore, the significant genetic differentiation observed between Tri1A and most of the samples may rather be due to a lower genetic diversity in this sample as compared to all others (Table 2, Chapuis and Estoup 2007) or a sampling bias associated with a Wahlund effect, also suggested by the observed heterozygote deficits (see above).

The fact that we observed only a slight significant genetic differentiation between samples located along the Italian coats, and no significant differentiation between the East and West Adriatic coasts suggests that *P. caerulea* forms here a large unique population. This pattern further suggests that *P. caerulea* planktonic larvae allow enough dispersion to cause genetic homogeneity across the study area. *P. caerulea* may therefore differ in life history traits compared to other limpets, e.g. *P. vulgata*, *P. candei*, *P. rustica* (Côrte-Real et al. 1996; Sá-Pinto et al. 2008) for which structured genetic variation has been observed at similar spatial scales. An alternative explanation could be related to the geological history of the Adriatic Sea. During the Last Glacial Maximum (about 18.000ya) the sea level was about 100 m below the actual mean water level, and most of the Adriatic Sea bed was dried (Dondi et al. 1985; Thiede 1978). The sea water invaded the Adriatic during the last 10.000 years and the colonization by the marine flora and fauna is very recent. Genetic similarities in Adriatic samples of *P. caerulea* may reflect past founder effects linked with the colonization of the Adriatic Sea after the Pleistocene glaciation. Indeed, several studies have recently stressed the relevance of palaeoecological events in determining the genetic patterns in marine populations (e.g. Fauvelot et al. 2003; Imron et al. 2007; Virgilio et al. 2009; Wilson 2006). Consequently, observed genetic patterns of *P. caerulea* in the Adriatic Sea likely reflect the interaction between historical events (long-term barriers followed by range expansion associated with Pleistocene sea level changes) and contemporary processes (gene flow modulated by life history and oceanography).

*Genetic diversity of* P. caerulea *populations on artificial and natural substrates*

One of the main outcomes of our study was the lower genetic diversity in populations from artificial structures compared to those from natural habitats. Indication of important effects of artificial substrata on the genetic structure of this limpet also comes from a previous study of Mauro et al. (2001), which found significant differences in the genetic structures of *P. caerulea* between artificial structures and natural rocky shores at two enzymatic systems out of twelve under study (*AAT** and *SOD-1**), though no differences in genetic diversity were observed among samples.

Altered genetic patterns and diversity may be expected in small, isolated, recently founded populations (Bradshaw et al. 2007; McElroy et al. 2003; Spencer et al. 2000) or in small founding populations of introduced species (Allendorf and Lundquist 2003; but see Roman and Darling 2007). However, we did not find evidence of recent bottlenecks in populations sampled on artificial substrates and *P. caerulea* is a native species in the study area. A related study on the gastropods *Nucella lapillus* (Colson and Hughes 2004) did not show reduced genetic diversity in recently colonized/recolonized populations. This discrepancy between studies could be related to differences in life-history traits between *P. caerulea* and *N. lapillus*, including differences in dispersal abilities, invasiveness, population turnover, and/or reproductive success (Johnson and Black 1984). Further, *N. lapillus* was sampled in natural habitats solely. Therefore, the lower genetic diversity observed in *P. caerulea* from artificial structures could also be related to the impact of artificial urban structures themselves. Indeed, in the Adriatic Sea, as well as in other geographical regions, there is growing evidence that artificial structures support assemblages that differ significantly in composition, structure, reproductive output, patterns of recruitment and population dynamics from assemblages on nearby natural rocky habitats (e.g. Bulleri 2005; Bulleri and Chapman 2004; Glasby and Connell 1999; Moschella et al 2005, Perkol-Finkel et al. 2006). These findings suggest important functional and ecological differences between these two types of habitats. For example, in Sydney Harbour (Australia) it has been shown experimentally that the reproductive output of populations of the limpet *Siphonaria denticulata* was significantly smaller on seawalls compared to natural shores, with possible important implications for the self-sustainability of local populations (Moreira et al. 2006). Also, variations in competition interactions on rocky shores and artificial structures have been observed among Mediterranean limpets (Espinosa et al. 2006). All these processes may act on propagule pressure (Lockwood et al. 2005) through small inoculum size (i.e the number of viable settlers), creating a filter from the amount of genetic diversity found in source populations, further causing genetic diversity to decrease, but maintaining genetic homogeneity between newly colonized and source populations (Roman and Darling 2007).

Urban structures and other artificial substrata are often uncritically claimed as reasonable mimics of natural hard-bottom habitats and valuable replacements for the habitats that they damage. Our results contribute to the growing body of evidence showing that although artificial structures attract and support species typical of hard bottoms, they are not analogues of natural rocky habitats (see among others Bulleri 2005; Glasby & Connell 1999; Moreira et al. 2006; Moschella et al. 2005). They can alter not only the identity and nature of marine coastal landscapes and the distribution of species, but also the genetic diversity of populations at local to regional scales. This is particularly important because the management of sea walls and similar artificial structures is generally carried out at local scales, without careful consideration of possible effects at larger spatial scales (Airoldi et al. 2005a). Future work should attempt to characterize more deeply how the type, quality and spatial arrangement (e.g. location relative to natural habitats and other artificial habitats) of fragmented artificial urban substrates affect the dispersal, distribution and genetic structure of species at a regional landscape scale, and the implications of these changes on the functioning of coastal marine systems at all spatial scales.

## ACKNOWLEDGMENTS

This work was supported by the EU projects DELOS (EVK3-CT-2000-00041) and EUMAR (EVK3CT2001-00048) and by the project MedRed (Bilateral Projects Italy-Israel, Italian Ministry of the Environment). We thank F. Bacchiocchi, C. Papetti, and M. Virgilio for help at various stages of the work. We are grateful to M. Coleman and anonymous reviewers whose comments greatly improved this manuscript.

**FIGURE LEGENDS**

**Figure 1**: Aerial view of the urban structures along the coasts of the Adriatic Sea (photo by Benelli, reproduced from Airoldi & Beck 2007, with permission).

**Figure 2**: Location of sampling areas of *Patella caerulea* in the Adriatic Sea. Within each of the five sampling area, at least two sites were sampled and when possible, both artificial and natural reefs were sampled in each site. Solid line: natural rocky coast; dash line: sandy coasts with hard artificial structures. N: natural habitat, A: artificial habitat

**Figure 3**: Mean allelic richness per locus (*Ar*) based on five analysed microsatellite loci (A) within each of the 14 sample sites based on 12 diploid individuals, white bars for artificial substrates, grey bars for natural shores, (B) for all 14 sample sites, average allelic richness per locus on artificial structures and natural shores based on 12 diploid individuals, and (C) only for direct pairwise comparisons (i.e. comparing only 8 artificial and natural shore sample sites : Tri1, Tri2, Ort1 and Ort2) and based on 22 diploid individuals.

**Table 1**: Primer sequences, repeat motif and amplification details for the five microsatellite loci specifically developed for *Patella caerulea*. Concentrations of $MgCl_2$ are given in mM.

| Locus | Accession no. | Primer sequences (5'-3') | Repeat motif | $MgCl_2$ | Cycles |
|---|---|---|---|---|---|
| *Pc11* | AY727872 | F : TTACGAAGCCCCAACTTCAC | $(AC)_3GC(AC)_7$ | 1.5 | 30 |
| | | R : AAGCCAGGGATAATGACACG | | | |
| *Pc15* | AY727873 | F : CCTTCTTCATGGGGACTTCA | $(TG)_{12}(TATG)_4(TG)_{12}$ | 1.5 | 30 |
| | | R : GCCCCAAAAACAATAGGGAT | | | |
| *Pc36* | AY727874 | F : GAACTAGCCGTGCCAATATGAT | $(CT)_{16}$ | 1 | 28 |
| | | R : GGTCGCTTCTGAGAAATGAAAT | | | |
| *Pc38* | AY727875 | F : GCTAATCTTTCAACGTATTTTT | $(AG)_{18}(AC)_6$ | 1.5 | 30 |
| | | R : GGTGTGGCTTGGAGATA | | | |
| *Pc73* | AY727876 | F : TGAAACAATATTCGCTGCTAGG | $(AC)_{11}CA(AC)_3CC(AC)_5$ | 1 | 27 |
| | | R : GCCCCAACGTAAAAATAACAGA | | | |

**Table 2**: Genetic diversity within *Patella caerulea* samples. *n*: total number of individuals genotyped, *N* : number of genotypes per locus ; *Ar*: allelic richness per locus and mean allelic richness per locus computed over all loci; $H_S$: gene diversity (Nei 1987); $H_O$: observed heterozygosity; *R*: null alleles frequency (Demspter estimator); significant deviations from Hardy-Weinberg equilibrium are indicated by an asterix following the $H_O$.

| | **Trieste** | | | | **Cesenatico** | | **Ortona** | | | | **Split** | | **Gallipoli** | |
|---|---|---|---|---|---|---|---|---|---|---|---|---|---|---|
| | **Tri1A** | **Tri1N** | **Tri2A** | **Tri2N** | **Ces1A** | **Ces2A** | **Ort1A** | **Ort1N** | **Ort2A** | **Ort2N** | **Spl1N** | **Spl2N** | **Gal1N** | **Gal2N** |
| *n* | 40 | 40 | 40 | 36 | 50 | 40 | 48 | 44 | 47 | 48 | 22 | 48 | 21 | 25 |
| ***Pc11*** | | | | | | | | | | | | | | |
| *N* | 40 | 40 | 39 | 34 | 50 | 40 | 48 | 44 | 47 | 48 | 22 | 48 | 21 | 25 |
| *Ar* | 5.80 | 5.61 | 6.40 | 6.48 | 6.32 | 6.64 | 6.21 | 7.42 | 6.44 | 6.95 | 4.53 | 6.62 | 5.62 | 7.23 |
| $H_S$ | 0.74 | 0.76 | 0.79 | 0.76 | 0.72 | 0.76 | 0.76 | 0.81 | 0.75 | 0.79 | 0.72 | 0.82 | 0.70 | 0.79 |
| $H_O$ | 0.58 | 0.68 | 0.72 | 0.56 | 0.60 | 0.65 | 0.77 | 0.75 | 0.79 | 0.83 | 0.64 | 0.63* | 0.48 | 0.52* |
| *R* | 0.08 | 0.03 | 0.08 | 0.19 | 0.05 | 0.06 | 0.00 | 0.02 | 0.00 | 0.00 | 0.07 | 0.12 | 0.12 | 0.15 |
| ***Pc15*** | | | | | | | | | | | | | | |
| *N* | 38 | 34 | 34 | 27 | 47 | 34 | 39 | 37 | 37 | 38 | 18 | 44 | 16 | 19 |
| *Ar* | 5.60 | 8.31 | 6.30 | 6.40 | 6.77 | 6.27 | 6.79 | 7.26 | 6.13 | 7.27 | 5.76 | 7.54 | 8.56 | 5.93 |
| $H_S$ | 0.51 | 0.78 | 0.77 | 0.78 | 0.70 | 0.69 | 0.74 | 0.78 | 0.76 | 0.80 | 0.74 | 0.81 | 0.83 | 0.79 |
| $H_O$ | 0.13* | 0.21* | 0.21* | 0.26* | 0.26* | 0.26* | 0.28* | 0.27* | 0.35* | 0.29* | 0.11* | 0.39* | 0.31* | 0.21* |
| *R* | 0.33 | 0.45 | 0.44 | 0.50 | 0.33 | 0.41 | 0.44 | 0.43 | 0.43 | 0.46 | 0.50 | 0.32 | 0.48 | 0.51 |
| ***Pc36*** | | | | | | | | | | | | | | |
| *N* | 29 | 35 | 22 | 29 | 35 | 32 | 36 | 36 | 33 | 39 | 12 | 31 | 20 | 17 |
| *Ar* | 9.39 | 11.23 | 11.28 | 10.05 | 11.54 | 9.87 | 10.08 | 10.99 | 9.57 | 10.07 | 10.00 | 10.72 | 8.96 | 9.78 |
| $H_S$ | 0.86 | 0.90 | 0.90 | 0.89 | 0.90 | 0.85 | 0.89 | 0.89 | 0.88 | 0.87 | 0.84 | 0.90 | 0.85 | 0.86 |
| $H_O$ | 0.38 | 0.17* | 0.23* | 0.28* | 0.26* | 0.28* | 0.25* | 0.42* | 0.36* | 0.36* | 0.42* | 0.13* | 0.30* | 0.18* |
| *R* | 0.48 | 0.47 | 0.66 | 0.47 | 0.55 | 0.47 | 0.52 | 0.40 | 0.51 | 0.43 | 0.61 | 0.63 | 0.34 | 0.59 |
| ***Pc38*** | | | | | | | | | | | | | | |
| *N* | 40 | 40 | 39 | 34 | 49 | 40 | 46 | 44 | 46 | 48 | 22 | 48 | 21 | 23 |
| *Ar* | 11.16 | 12.41 | 11.02 | 10.47 | 10.50 | 10.46 | 11.40 | 11.46 | 11.66 | 11.95 | 13.04 | 12.24 | 14.65 | 13.31 |
| $H_S$ | 0.87 | 0.90 | 0.84 | 0.85 | 0.89 | 0.87 | 0.88 | 0.89 | 0.90 | 0.90 | 0.91 | 0.91 | 0.92 | 0.89 |
| $H_O$ | 0.85 | 0.80 | 0.72 | 0.94 | 0.73* | 0.75 | 0.70* | 0.84 | 0.83 | 0.79 | 0.82 | 0.79 | 0.86 | 0.65* |
| *R* | 0.03 | 0.05 | 0.10 | 0.07 | 0.11 | 0.07 | 0.14 | 0.04 | 0.07 | 0.05 | 0.02 | 0.06 | 0.03 | 0.21 |
| ***Pc73*** | | | | | | | | | | | | | | |
| *N* | 32 | 34 | 34 | 32 | 38 | 35 | 42 | 40 | 39 | 45 | 17 | 31 | 19 | 18 |
| *Ar* | 8.52 | 7.74 | 7.99 | 9.91 | 8.56 | 7.46 | 7.65 | 9.99 | 8.28 | 7.68 | 10.14 | 8.24 | 8.94 | 7.64 |
| $H_S$ | 0.81 | 0.84 | 0.79 | 0.87 | 0.83 | 0.83 | 0.83 | 0.87 | 0.84 | 0.82 | 0.89 | 0.84 | 0.86 | 0.81 |
| $H_O$ | 0.06* | 0.12* | 0.24* | 0.25* | 0.11* | 0.17* | 0.36* | 0.15* | 0.21* | 0.18* | 0.18* | 0.13* | 0.16* | 0.06* |
| *R* | 0.55 | 0.50 | 0.44 | 0.42 | 0.56 | 0.45 | 0.37 | 0.45 | 0.47 | 0.40 | 0.53 | 0.62 | 0.45 | 0.60 |
| ***Multilocus*** | | | | | | | | | | | | | | |
| *Ar* | 8.09 | 9.06 | 8.60 | 8.66 | 8.74 | 8.14 | 8.42 | 9.42 | 8.42 | 8.79 | 8.69 | 9.07 | 9.35 | 8.78 |
| $H_S$ | 0.76 | 0.84 | 0.82 | 0.83 | 0.81 | 0.80 | 0.82 | 0.85 | 0.83 | 0.83 | 0.82 | 0.86 | 0.83 | 0.83 |
| $H_O$ | 0.40* | 0.39* | 0.42* | 0.46* | 0.39* | 0.42* | 0.47* | 0.49* | 0.51* | 0.49* | 0.43* | 0.41* | 0.42* | 0.32* |

**Table 3**: Genetic differentiation of *Patella caerulea* among 14 sample sites obtained from the analysis of five microsatellite loci. Pairwise $F_{ST}$ estimates (Weir 1996) computed following the ENA method (Chapuis and Estoup 2007). Values in italics indicate significant genotypic differentiation of the samples at the 5% threshold and those in bold indicate significant genotypic differentiation of the samples after sequential Bonferroni correction of the 5% threshold.

| | **Tri1N** | **Tri2A** | **Tri2N** | **Ces1A** | **Ces2A** | **Ort1A** | **Ort1N** | **Ort2A** | **Ort2N** | **Spl1N** | **Spl2N** | **Gal1N** | **Gal2N** |
|---|---|---|---|---|---|---|---|---|---|---|---|---|---|
| **Tri1A** | 0.019 | **0.031** | **0.029** | **0.013** | **0.014** | **0.020** | **0.024** | **0.019** | **0.021** | *0.025* | **0.029** | **0.032** | 0.022 |
| **Tri1N** | | 0.006 | **0.013** | *0.007* | 0.007 | **0.011** | 0.004 | 0.006 | 0.009 | -0.005 | *0.015* | 0.008 | 0.003 |
| **Tri2A** | | | *0.014* | **0.025** | *0.015* | **0.021** | *0.010* | 0.011 | *0.014* | 0.010 | 0.017 | 0.013 | 0.000 |
| **Tri2N** | | | | *0.012* | 0.004 | 0.004 | 0.004 | *0.007* | *0.009* | *0.012* | 0.007 | 0.004 | 0.004 |
| **Ces1A** | | | | | 0.000 | 0.003 | 0.007 | 0.004 | 0.005 | *0.013* | **0.017** | 0.008 | 0.010 |
| **Ces2A** | | | | | | -0.002 | 0.002 | 0.000 | 0.003 | 0.012 | 0.011 | 0.002 | 0.007 |
| **Ort1A** | | | | | | | 0.003 | 0.004 | 0.003 | **0.016** | 0.010 | 0.004 | 0.007 |
| **Ort1N** | | | | | | | | 0.000 | 0.001 | 0.005 | *0.007* | 0.004 | 0.003 |
| **Ort2A** | | | | | | | | | -0.002 | *0.007* | 0.010 | -0.001 | 0.002 |
| **Ort2N** | | | | | | | | | | *0.013* | 0.011 | 0.002 | 0.002 |
| **Spl1N** | | | | | | | | | | | 0.009 | 0.011 | 0.005 |
| **Spl2N** | | | | | | | | | | | | 0.007 | 0.005 |
| **Gal1N** | | | | | | | | | | | | | -0.001 |

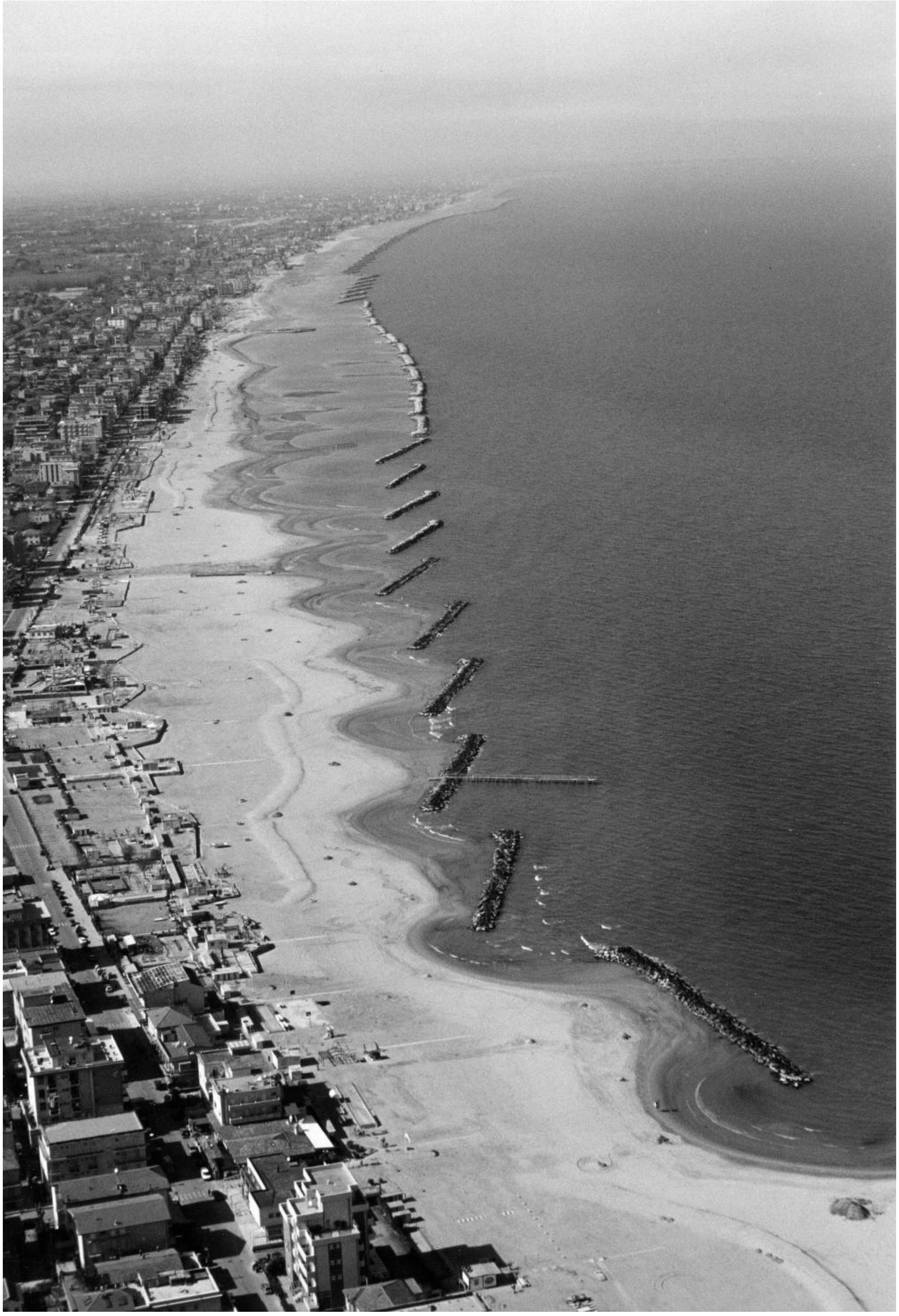

**Figure 1**: Aerial view of the urban structures along the coasts of the Adriatic Sea (photo by Benelli, reproduced from Airoldi & Beck 2007, with permission).

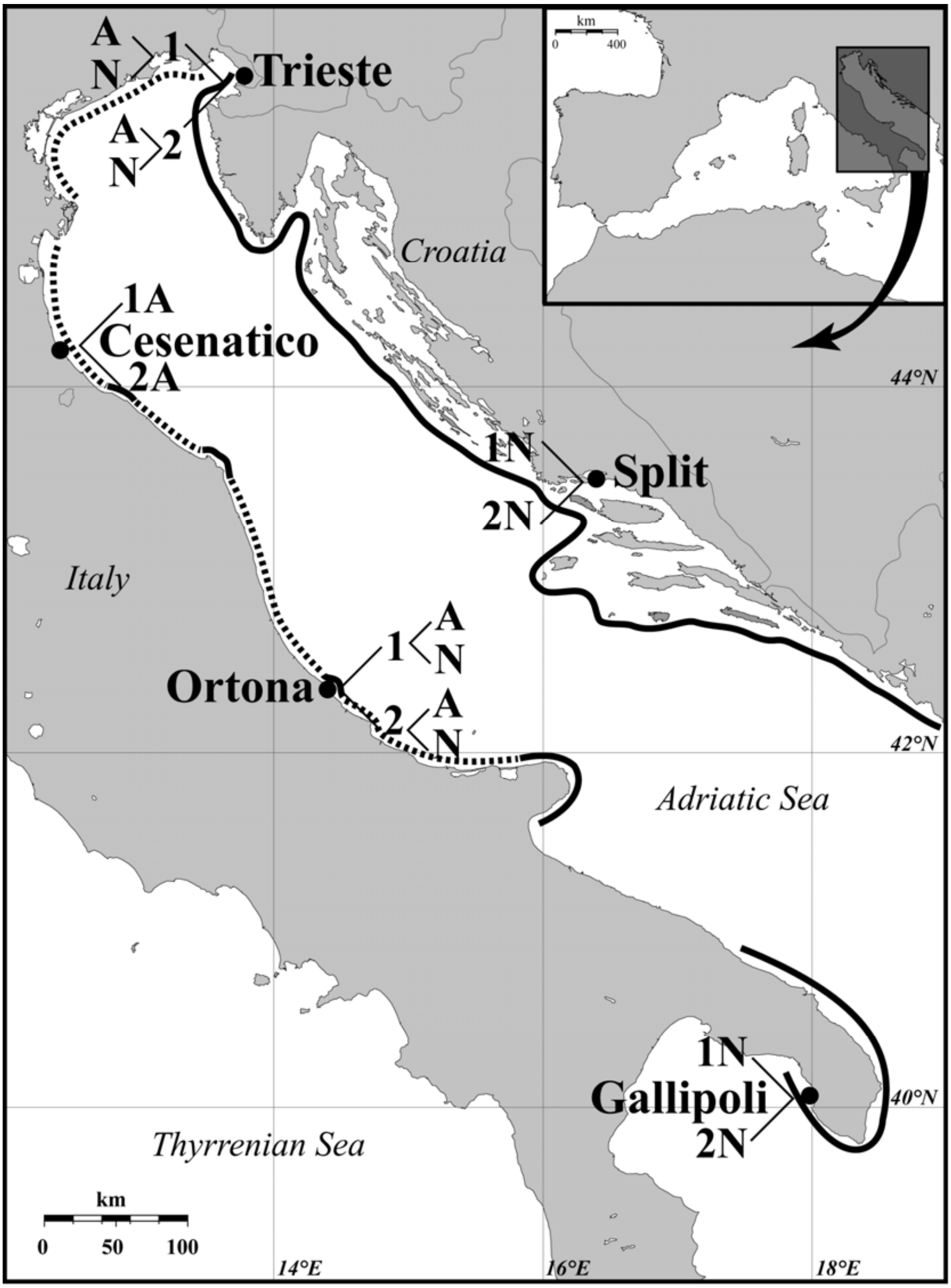

**Figure 2**: Location of sampling areas of *Patella caerulea* in the Adriatic Sea. Within each of the five sampling area, at least two sites were sampled and when possible, both artificial and natural reefs were sampled in each site. Solid line: natural rocky coast; dash line: sandy coasts with hard artificial structures. N: natural habitat, A: artificial habitat

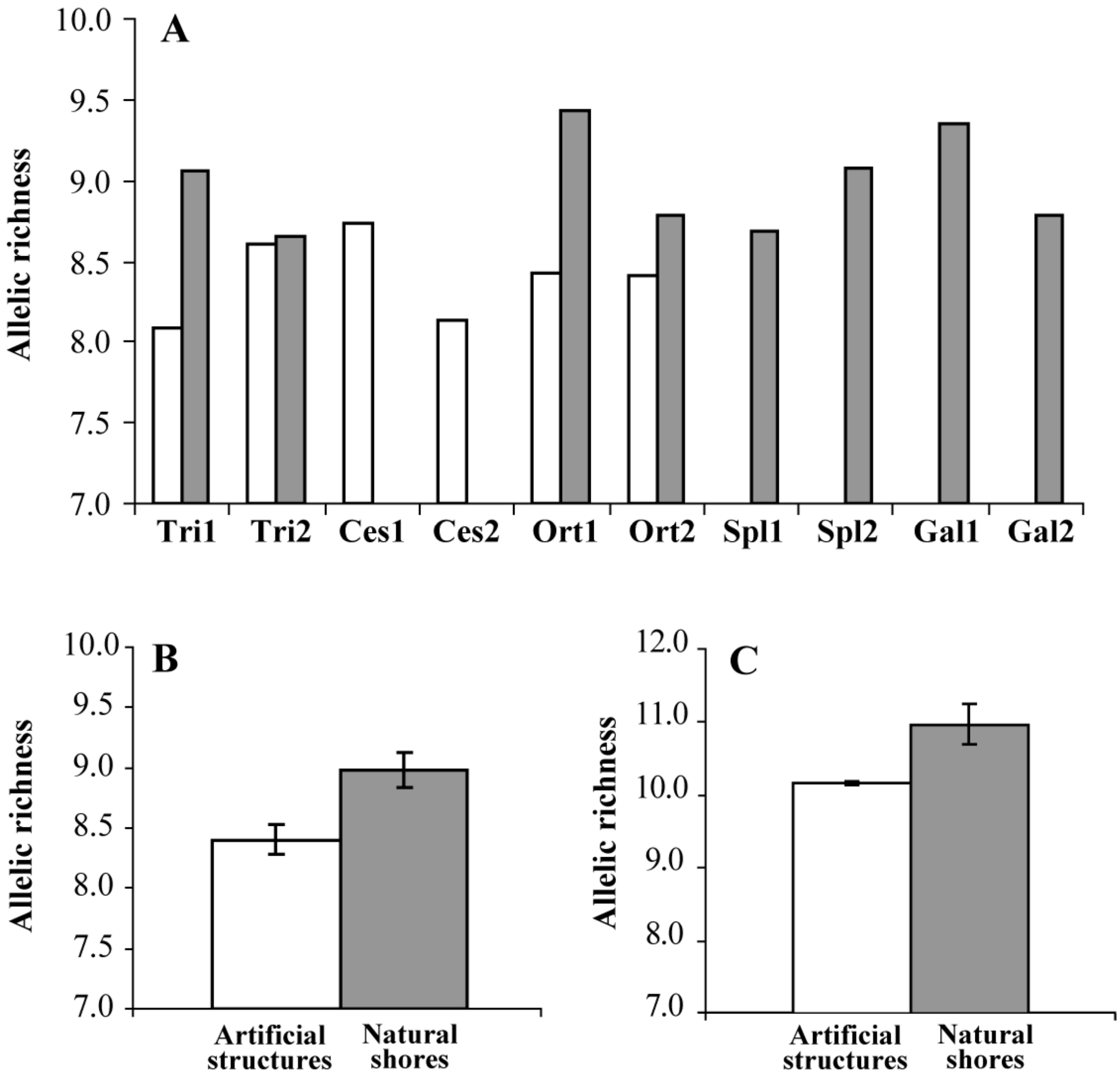

**Figure 3**: Mean allelic richness per locus (*Ar*) based on five analysed microsatellite loci (A) within each of the 14 sample sites based on 12 diploid individuals, white bars for artificial substrates, grey bars for natural shores, (B) for all 14 sample sites, average allelic richness per locus on artificial structures and natural shores based on 12 diploid individuals, and (C) only for direct pairwise comparisons (i.e. comparing only 8 artificial and natural shore sample sites : Tri1, Tri2, Ort1 and Ort2) and based on 22 diploid individuals.